\documentclass[twocolumn,showpacs,preprintnumbers,amsmath,amssymb]{revtex4}
\usepackage{graphicx}
\usepackage{dcolumn}
\usepackage{bm}

\begin{document}

\preprint{cond-mat/0405363}

\title{Propagation of Ripples in Monte Carlo Models \\ of Sputter
  Induced Surface Morphology}

\author{Emmanuel O. Yewande}
\email{yewande@theorie.physik.uni-goettingen.de}
\author{Alexander K. Hartmann}
\author{Reiner Kree}
\affiliation{Institut f\"ur Theoretische Physik,
   Universit\"at G\"ottingen, Friedrich-Hund Platz 1, 37077
   G\"ottingen, Germany 
}

\date{\today}
\begin{abstract}
Periodic ripples generated from the off-normal incidence ion beam bombardment
  of solid surfaces have been observed to propagate with a dispersion in the 
  velocity. We investigate this ripple behaviour by means of a Monte Carlo
  model of the erosion process, in 
  conjuction with one of two different surface diffusion mechanisms, 
representative of two different classes of materials; one is a
Arrhenius-type Monte Carlo method including a term (possibly zero)  that
accounts for the Schwoebel effect, the other
a thermodynamic mechanism without the Schwoebel effect. We
  find that the behavior of the ripple velocity and wavelength depends
  on the sputtering timescale, qualitatively consistent with
  experiments. Futhermore, we observe a strong temperature dependance
  of the ripple velocity, calling for experiments at different temperatures.
 Also, we observe that the ripple velocity
 vanishes ahead of the periodic ripple pattern.  
\end{abstract}

\maketitle
\section{\label{sec:intro}INTRODUCTION}
There has been much scientific activity for quite some time now, on the 
features of surface
morphology resulting from the bombardment of a solid surface by a collimated
beam of intermediate energy ions, at normal and oblique incidence to the solid 
surface \cite{r1, makeev2002}. The phenomenon is an 
essential constituent of several surface
analysis, processing and fabrication techniques, such as ion beam aided
deposition, surface catalysis, 
sputter cleaning, etching and deposition. 

Normal incidence ion bombardment of
non-metallic substrates often results in an interlocking grid of hillocks
and depressions, which have been demonstrated to be an attractive
alternative to the spontaneous growth of self-organized quantum dots
on semiconductor surfaces in the
Stranski-Krastanov growth mode \cite{r3}. Off-normal incidence ion
bombardment of such non-metallic substrates, however, gives rise to 
the  formation of quasiperiodic ripples \cite{r4, r5, r6, r7, r8, carter1996,
  r10, r11} with   
orientation that depends on the 
angle of incidence of the ion beam. For incidence angles less than a critical
angle, $\theta_c$ \cite{r12}, the wavevector of the ripples is parallel
to the projection 
of the ion beam direction on the surface plane while for incidence angles
greater 
than $\theta_c$, the wavevector of these ripples is oriented perpendicular  to
the  projection of
the ion beam direction on the surface plane. On the other hand, ripples are
observed on metallic substrates at normal incidence ion bombardment, and these
ripples are rotated by changing the subtrate temperature \cite{r13, r14, r15};
a probable consequence of the symmetry-breaking anisotropy in surface
diffusion. The wavelengths of the observed ripples, in all cases, is
of the order of 
tenths of micrometers.

However, a number of experimental studies \cite{r16, r17, r18, r19} have
demonstrated that under certain ion 
bombardment conditions, ripples are not formed; the surface undergoes kinetic
roughening with interesting scaling properties. All these observations point
to the possibility of several phases in the surface topography evolution, with
phase boundaries defined by the bombardment conditions, and with little or no
dependence on the material composition, surface chemistry, defects or chemical
reactions on the surface.
These features are understood, from insightful theoretical descriptions
 \cite{makeev2002, bradley1988, Cuerno1995}, as being governed by the interplay and
 competition between 
the dynamics of surface roughening on the one hand and material transport
during surface migration on the other. 
Ion bombardment tends to roughen the surface, while surface 
diffusion leads, in general, to surface relaxation \cite{r4,r5}
For sufficiently low ion energies, the
sputtering phenomenon is the dominating mechanism \cite{makeev2002}.
However, if the flux is low at such energies, then 
the enhanced defect mobility can result to domination by surface
diffusion  which may 
cause the overall scaling behavior of the surface profile to be uniquely 
determined by the nonequilibrium biased diffusion current, independently 
of the microscopic origin \cite{pimpinelli2002}. 

Recently surface ripples generated during Gallium ion beam erosion of
Silicon were observed to propagate with a ripple velocity that scales
with the ripple
wavelength as $v\sim \lambda^{k}$, where $k\simeq 0$ initially, and
$k=-1.5$ after a crossover wavelength $\lambda_c\simeq 100$nm 
\cite{habenicht2002}.  
 This velocity dispersion has
 been ascribed to an indication of a continuous transition to a rising
 non-linear contribution in surface erosion \cite{makeev2002,
   habenicht2002}. Motivated  by this experimental result, we study ripple 
propagation by means of a recently introduced, discrete (2+1)
dimensional Monte Carlo (MC) model \cite{hartmann2002} of the sputtering
process, and 
two different solid on solid models of surface diffusion; for details see
below . We focus on intermediate times, where the transition from
linear to non-linear regimes occur. Our results corroborate the experimental
observation, but in addition, we find that 
, at high temperatures,
the
ripples first come to rest before they are 
completely wiped out by the increasing non-linear contributions.    

The rest of the paper is organized as follows. First, we state our
simulation model, i.e. how the sputtering process and the different
diffusion mechanisms are implemented. Then we explain, how we study the
movement of the ripples. In the main section, we show our simulation
results. We finish with our conclusions and an outlook.

\section{\label{sec:algorithm} Erosion and surface migration}

According to Sigmund's sputtering theory \cite{r24}, the rate at which
material is 
removed from a solid surface, through the impact of energetic particles, is
 proportional to the power deposited there by the random slowing down of
 particles. The average energy $E({\bf r}^\prime)$ deposited at surface point
 ${\bf r}^\prime=(x^\prime, y^\prime, -z^\prime)$ is given by the Gaussian
 distribution 
\begin{equation}
E({\bf r}^\prime)= \frac{\epsilon}{(2\pi )^{3/2}\sigma\mu^2}\exp\biggl (
-\frac{(z^\prime+d)^2}{2\sigma^2}-\frac{x^{\prime^2}+y^{\prime^2}}{2\mu^2}
\biggr ) 
\label{eq:sigmund}
\end{equation}
where we have used the local Cartesian coordinate system of the ion with
origin at the point of penetration and with the $z$ axis coinciding with the
ion beam direction; $(z^\prime+d)$ is the distance of the surface
point, from final stopping point of ion, measured along the ion
trajectory, $\sqrt{x^{\prime^2}+y^{\prime^2}}$ is the 
distance perpendicular to it; $\sigma$ and $\mu$ are the widths of the
distribution parallel and perpendicular to the ion trajectory respectively;
$\epsilon$ is the total energy deposited, $d$ is the average depth of energy
deposition. 
Sigmund's formula is the basis for {\em all} theoretical treatments
and analysis of experimental results so far.

\subsection{\label{sec:sputtering}The Sputtering Process}
Following \cite{hartmann2002}, 
we simulate the sputtering process on a surface of size $L^2$ with periodic
boundary conditions, by starting an ion at
a random position in a plane parallel to the plane of the initially flat
surface, and projecting it along a straight trajectory inclined at
angle $\theta$ to 
the normal to the average surface configuration; at an azimuthal angle
$\phi$. The ion penetrates the solid through a
depth $d$ and releases its energy, such that an atom at a position ${\bf r}=(x,
y, h)$ is eroded (see Fig. 1 of \cite{hartmann2002}) with probability
proportional to $E({\bf r})$. It should be noted that,
consistent with the assumptions of the theoretical models 
\cite{makeev2002, bradley1988, Cuerno1995}, this
sputtering model assumes no evaporation, no redeposition of eroded
material, no preferential sputtering of surface material at point of
penetration, and surface is defined by a single valued, discrete time
dependent height function $h(x, y, t)$ (solid-on-solid model, SOS). 
The time t is measured in terms of the
ion fluence; i.e, number of incident ions per two-dimensional 
lattice site $(x,y)$. We used incidence
angle $\theta = 50^\circ$, azimuthal angle $\phi = 22.0^\circ$, $d=6.0$,
$\sigma = 3.0$, $\mu = 1.5$, 
as obtained by SRIM \cite{srim} for 5keV Xe$^+$ ions on
graphite and rescaling all lengths by a factor 2.
This should give 
according to the linear theory of Bradley
and Harper a value $\theta_c = 68^\circ$ \cite{r12}. We have chosen
$\epsilon$  to be $(2\pi)^{3/2}\sigma \mu^2$, which leads to high
sputtering yields $Y\simeq 7.0$ compared to experiments like
\cite{r11}, where $Y=0.3,..., 0.5$, i.e. increasing the efficiency of
the simulation. According to the Bradley Harper
theory, the ripple wavelength $\lambda$ scales like $\lambda\sim
Y^{-1/2}$ so that we expect patterns with
correspondingly smaller length scales in our simulations. This we have
to remember when quantitatively interpreting the result. Anyway, the
general phenomena observed in the simulation are not affected by this
choice.

Our model of the sputtering mechanism sets the time scale of the
simulation, and allows comparison with experiments. Additionally, also
moves of atoms mimicking surface diffusion are performed, described now.

\subsection{\label{sec:diffusion}The Hamiltonian and Arrhenius Models of
  Surface Diffusion}
Surface migration is modelled as a thermally activated nearest neighbor
hopping process, as in \cite{Siegert1994, Smilauer1993}. 
A Monte Carlo acceptance/rejection procedure is used for this purpose.
One diffusion step refers to a
complete sweep of the lattice. Two different solid-on-solid models of
surface diffusion in molecular beam epitaxy are used; the second one
 of them sensitive
to the repulsion of a diffusing particle from a down step, and preferential
diffusion in the uphill direction: the so-called Schwoebel effect. 

The first model \cite{Siegert1994} is based on a thermodynamic interpretation
of the diffusion process. For each step,
a site $i$ and one neighbour site $j$
are randomly selected. The trial move is an atom hopping from $i$ to
$j$, i.e. $h_i=h_i-1$ and $h_j=h_j+1$.
We calculate the surface energy before
and after the hop, 
through the energy of an unrestricted SOS model
 \begin{equation}   
E=\frac{J}{2}\sum_{<i, j>}|h_i-h_j|^2
 \end{equation}  
$J$ is a coupling constant through which the nearest neighbor sites
 interact. $h_i$ is the height variable at site $i$, 
and the summation is over the nearest neighbors on the
 2-dimensional substrate. 

The hop is allowed with the  
probability
\begin{equation}
p_{i\rightarrow f}=1/\biggl [1+\exp \biggl (\frac{\triangle E_{i\rightarrow
      f}}{k_BT}\biggr )\biggr ]
\label{eq:hop-prob}
\end{equation}
where $\triangle E_{i\rightarrow f}$ is the energy difference between the
  initial and final states of the move. $T$ is the substrate
  temperature, and $k_B$ is the Boltzmann's constant.  
Although no exact mapping is
  possible we can estimate that a temperature $k_bT/J 
\simeq 0.2$
  in this model corresponds roughly to the 
temperature used in the second model below.
  The estimate
  is based on a comparison of the pure diffusion mechanism without
  sputtering such that
  they lead to comparable values of the roughness.
Note that 
this 
temperature is below
  the roughening transition of this model \cite{Siegert1994}.
This model does not prevent atoms from
 moving down over steps edges, hence no Schwoebel effect is present.

The second model is also based on a MC procedure and uses
a formula known from kinetic MC mechanisms.
For each step, again a site $i$ and a nearest neighbor site
$j$ are  chosen at random but now a
hopping move is performed with a probability proportional to the hopping
rate of an Arrhenius form 
\begin{equation}
k(E, T)=k_0 \exp\biggl ( -\frac{E}{k_BT}\biggr )
\label{eq:hopping-rate}
\end{equation}
 $E=E_{SB}+nnE_{NN}+E_S$ is an energy barrier to hopping, consisting of a
Schwoebel barrier term $E_{SB}$, a substrate term $E_S=0.75$eV and a nearest
neighbor bonding of magnitude $nnE_{NN}=nn0.18$eV;
 where $nn$ is the number of in-plane
nearest neighbors of the diffusing atom. $E_{SB}$ is equal to some
constant ($0.15$eV in this case. Note that we perform also runs for $E_{SB}=0$,
to compare with the thermodynamic model, see below), if
the numbers of next-nearest neighbors in the plane beneath the hopping atom
before ($nnn_b$) and after ($nnn_a$) the hop, obey $4 = nnn_b > nnn_a$; and
zero otherwise. 
Our temperature is measured in units of eV$k_B^{-1}$ in this model,
where $T \simeq 0.02eVk_B^{-1}$ corresponds to room temperature. 
$k_0=2k_BT/h$ is the vibrational frequency of a surface
adatom, i.e. a hopping attempt rate, $h$ being Planck's constant. The hopping
attempt rate is very high, with a corresponding low hopping
probability resulting from Eq. \ref{eq:hopping-rate}, 
slowing down the simulation. Thus we 
incorporate the factor
$\exp(-E_S/k_BT)$ into the rescaled attempt rate such that the hopping
rate reads 
\begin{equation}
k(E, T)=k_1 \exp\biggl ( -\frac{\triangle E}{k_BT}\biggr )
\end{equation}
where $k_1=k_0 \exp\biggl ( -\frac{E_S}{k_BT}\biggr )$ is a much
lower hopping attempt rate, $\triangle E=nnE_{NN}+E_{SB}$.  
This physical attempt rate, in comparison with the ion current density
used in experiments, determines the ratio between the number of 
sputtering steps
and the number of surface diffusion steps made in the simulation. In the next
section, we state the values we used for our simulations. A discussion
of parameter optimization and a rescaling of the temperature with the
parameters is given in \cite{Shitara1992} 
Note finally that
for atoms on top of planes, which are far from down edges, $\Delta E =0$,
i.e. each hop is accepted, independently of the temperature.

\section{\label{sec:kinematics}Ripple Kinematics}

In experiments, we typically have $N = 1$x$10^{15}$atoms/cm$^2$ on the
surface. Since
 the typical experimental ion current density is
of the order $F=7.5$x$10^{14}$ ions/cm$^2$s \cite{habenicht2002}, this
 implies a flux of $\Phi = F/N \simeq 0.75$ ion/atom s. 
From the values given above, we get hopping
attempt rates $k_1$ of around $200$
 1/s for room temperature, hence 200 sweeps of the diffusion mechanism
 correspond  to 0.75 ions per surface atom. 
Thus, we initiate a diffusion step  every $\Phi L^2/k_1 =
 0.0037L^2$ erosion steps; $L$ is the system size.      

\begin{figure}[hbtp!]
\includegraphics[angle=270, width=0.45\textwidth]{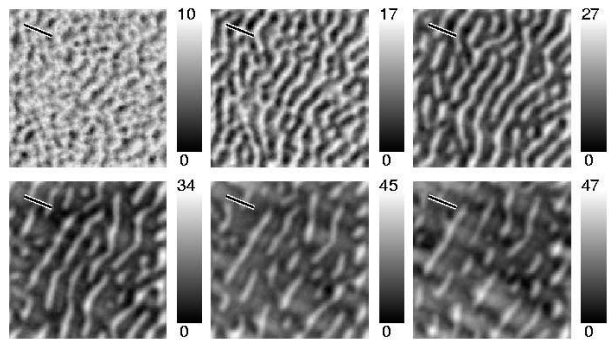}
\caption{\label{fig:profile1}Surface profiles at a substrate temperature
  of 0.2$Jk_B^{-1}$ and at different times. Starting from top-bottom, 
  left-right, t=0.5, 1.5, 4.0, 9.0, 14.0 and 20.0 ions/atom. Ion beam
  direction, indicated by the bar, is perpendicular to ripple
  orientation. The scales 
  show the surface height measured from the lowest height.}
\end{figure}

Initially, for times less than about $1.4$ ions/lattice site, the
surface is rough \cite{hartmann2002} and then the formation of  
ripples starts. In this paper we focus on 
the motion and time development of these ripples. In
Fig. \ref{fig:profile1} the time development of a 
sample surface topology is shown for
the first diffusion model. Initially ripples are formed. They
propagate slowly and, due to the increasing influence of non-linear
effects (note the scales at the right), disappear at longer times. 
 The long-time
behavior, where the ripples have disappeared, has already been studied
in Ref. \onlinecite{hartmann2002}.

In order to monitor the ripple propagation on the computer, 
we assign the crest points of the
ripples to clusters, and then monitor the motion of these clusters. 
A cluster of crest points is defined as the set of surface points with
height $h(x, y, t)\ge h_c$ and nearest neighbor distance $l \le l_c$, where
$h_c$ and $l_c$ are cut-off surface height and 
distance between neighboring cluster points respectively.  We
have chosen our cut-off
height to be a function of the average height of the configuration
$\langle h\rangle$, and
the height difference $h_d$ between the maxima and minima of the surface;
i.e, $h_c=\langle h\rangle + ph_d$, where $p$ is some fixed
percentage. In this way clusters with about the same proportionate
sizes can be followed from the beginning of ripple
formation until complete disappearance of the ripples. 
Furthermore, we have used $l_c=2$. Different,
unconnected ripples should, in general, generate different clusters. We also
require that the number $N$ of elements in a cluster be large
enough to allow for statistical analysis, here we have chosen $N\ge
10$ elements. 

\begin{figure}[hbtp!]
\includegraphics[width=0.23\textwidth]{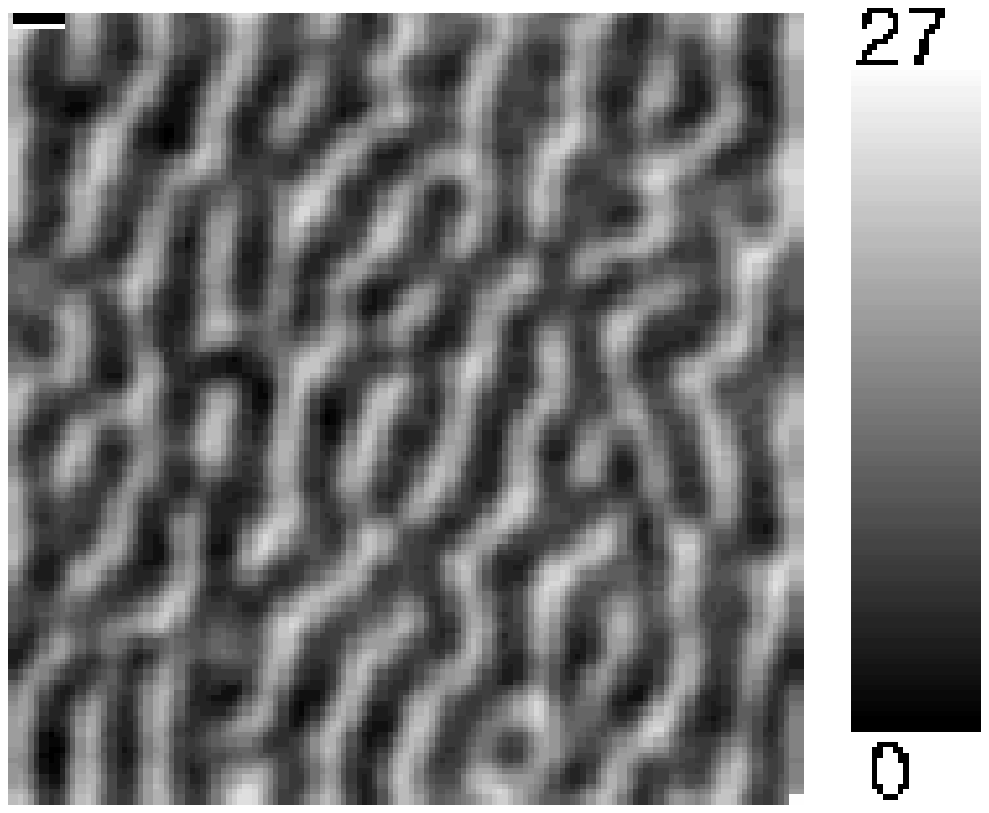}
\includegraphics[width=0.23\textwidth]{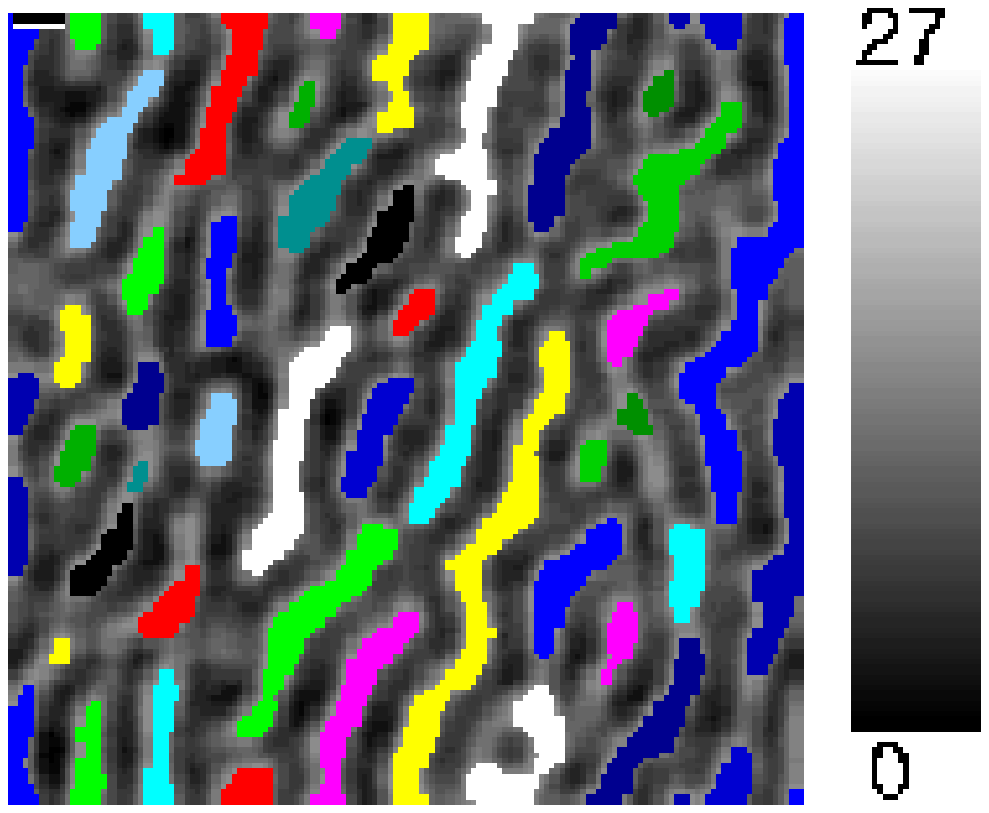}
\caption{\label{fig:kinematics}(Color online) Surface profile for 
time 3 ions/atom (thermodynamic diffusion model, 
$T=0.2J/k_B$, $L=128$). 
The second profile
  contains the clusters formed from the first profile, as described in
  the text.}
\end{figure}

The propagation of the ripples is studied
by calculating the time rate of change of the position of the centre
of mass of a cluster
\begin{equation}
\dot{\bf x}_{CM}=\frac{\sum_im_i\dot{\bf x}_i}{\sum_im_i}
\end{equation}
where the summation is over all the elements of the cluster. We have
assumed a homogeneous system composed of unit mass particles, such
that the center of mass of a cluster is ${\bf x}_{CM}=N^{-1}\sum_i{\bf
x}_i$. The ripple 
wavelength is given by $\lambda = 2\pi/\eta$, $\eta$ being the average
expectation value of the Gaussian fitted to the peak of the structure factor
$S({\bf k})=|h({\bf k})|^2$, where $h({\bf k})$ is the fourier
transform of the height topography $h({\bf r}, t)$, given by 
\begin{equation}
h({\bf k})=\frac{1}{L^{d^\prime/2}}\sum_{\bf r}[h({\bf r}, t)-\langle
h\rangle]e^{i{\bf kr}} 
\end{equation} 
$d^\prime$ is the substrate dimension, i.e. here $d^\prime=2$.
Fig. \ref{fig:kinematics} 
shows two profiles of the surface for system size $128$x$128$ at
time $t=3$ ions/atom; in the second profile, we print the clusters on top
of their corresponding ripples. As seen in the figure of the clusters,
application of periodic boundary conditions neccessitates the need to
first unfold toroidal clusters before calculating the position of
their center of mass. As time increases, local surface slopes
$\nabla h$ increase, and since the non-linear effects depend on the
square of $\nabla h$ they will dominate by scaling down surface
relaxation mechanisms \cite{r1}. These non-linear effects are responsible for
the disapearance of ripples (Fig. \ref{fig:profile1}) 
at long times, and for the transition of
the surface topography from a periodic ripple pattern to a rough
topography with self-affine scaling \cite{makeev2002, r16}. We thus expect 
fluctuations in the position of the centre of mass due to
disappearing ripples; the fluctuations are averaged out by using
systems of size $512$x$512$ with a large number of clusters such that
the ripple velocity at any time is an average of the velocities of all
the ripples at this time.

\section{\label{sec:results}Results and Discussion}

The results are obtained, as already mentioned, 
for square lattices of size $512$x$512$,
with periodic 
boundary conditions, and as an average over fifty different realizations.

\begin{figure}[!htbp]
\begin{center}
\includegraphics[angle=0, width=0.35\textwidth]{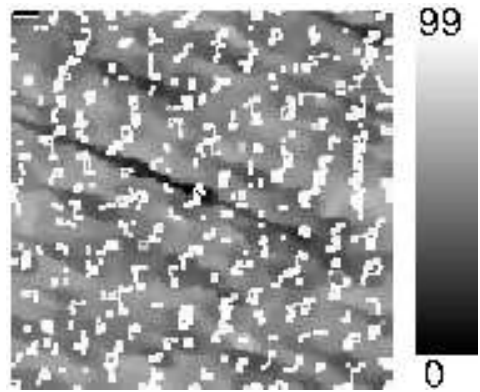}
\caption{\label{fig:schwoebel_room} Sample surface topology for a
  small system (L=128), for the Arrhenius MC diffusion mechanism at
  surface temperature equal to room temperature, 
after $t=100$ ions/atom. No clear ripples can be
  observed. Similar results were observed for almost all time, except
  the very early ones.}
\end{center}
\end{figure}

For the case of the Arrhenius diffusion mechanism, 
(including the Schwoebel barriers) one can choose a temperature
corresponding to the physical temperature present in the experimental system.
A naive guess
 is to use
room temperature $k_BT=0.02$eV, at which the experiments usually are
carried through. The resulting structures are 
shown in Fig. \ref{fig:schwoebel_room}, for intermediate as well after 
long sputtering times. We cannot observe clean ripples. 
The reason is that this kind of diffusion
mechanism is too slow at room temperature to effectively counteract
the strong roughening due to our model of sputtering, which possesses
a particularly high sputtering yield.  Hops are almost always
prevented if an atom has in-plane neighbors, so the mechanism is not
very effective on a rough surface. 
Since the surface relaxation is essential for the
formation of ripples \cite{bradley1988}, it needs 
{\em locally} higher than room temperatures
to produce clean ripples in our model. 
This happens indeed in experiments, since most of the
kinetic energy, carried by the incoming ion, is converted into lattice
vibrations, hence the surface is locally strongly heated. 
Here, we do
not know the spatio-temporal distribution of the local temperature.
Either one would have to perform MD simulations, or include heat
conduction in the model, both making the treatment of large systems
over long time scales infeasible. Instead, we are
choosing a higher but constant effective temperature $T$, which is a 
good first approximation.

Now, we want to estimate this effective temperature. The most basic approach
is to describe the energy carried by the ions as a constant inflow of
energy at the surface, fix the temperature far away from the surface
to room temperature and solve the stationary heat-conduction equation
to calculate the temperature at the surface \cite{melngailis1987}. The
resulting temperature depends strongly on the ion energy, the 
ion current density, and the
thermal conductivity of the material. For experimentally reported
parameters, temperature rises up to 1500 K ($0.155$eV$k_B^{-1}$) are found
\cite{melngailis1987}. This shows that high efficient temperatures,
even in the stationary state, may be achieved. However, in the
experiments of Habenicht et al. \cite{habenicht2002} only small
average ion
current densities have been used, which result in a temperature rise
at the surface of only few K. 

This does not mean that one can use a
temperature close to room temperature as effective temperature. The
reason is that right after impact, the surface is strongly heated
close to the melting temperature and
the quickly cooled again, i.e. a {\em thermal spike}
occurs \cite{primak1955}. Furthermore, the surface is sputtered using a focused
ion beam (of diameter 30nm), which 
is moved relatively slowly over the surface and
which exhibits a large spot current of 15$\mu$A/cm$^2$.
Hence, under the ion beam, for several short time intervals, 
surface diffusion is greatly enhanced. Marks has calculated \cite{marks1997}
the spatio-temporal development of the
temperature after ion impact by solving the dynamic heat-conduction
equation, resulting in a temperature profile $\tilde{T}(r,t)$ as function of
time $t$ and distance $r$ from the point of impact. The initial
distribution $\tilde{T}(r,0)$ is given by a step function with 
$\tilde{T}(r,0)$ being
the melting temperature of the material for $r\le r_0$ and being the room
temperature elsewhere. The initial radius $r_0$ is determined such that the
thermal energy inside this semi sphere equals to the energy carried by
the ion. Marks found that the surface is heated strongly right after
the impact and is cooled down to temperatures close to room
temperature within few ps.
\begin{figure}[hbtp!]
\includegraphics[angle=270, width=0.45\textwidth]{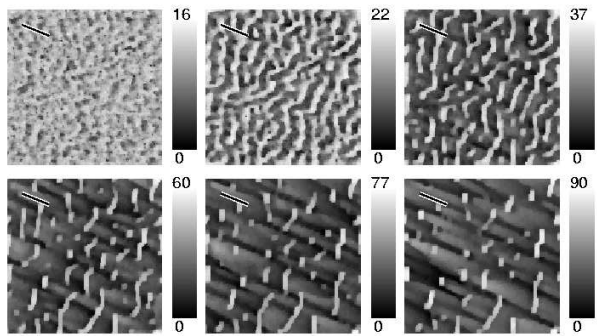}
\caption{\label{fig:profile2}Surface profiles at a substrate temperature
  of 0.1eV/$k_B$ with the second diffusion model. Starting from top-bottom, 
  left-right, t=0.5, 1.5, 4.0, 9.0, 14.0 and 20.0 ions/atom. In both
  cases, depicted here and in Fig. 1, ripples propagate along a
  direction opposite to that of the ion beam.}
\end{figure}
Qualitatively and quantitatively similar profiles have been observed
in MD simulations \cite{saada1999} as well. We apply
his equation, using the parameters for ion energy and ion current
density in the spot as given above, 
to determine an effective temperature (with $r_0=15.6\AA$ in our
case).  The basic idea is that in an time interval $\Delta t$, the
number of hops governd by
the temperature $\tilde{T}(0,t)$ at the impact point should be the same
as under the effective temperature $T$:
\begin{equation}
\int_0^{\Delta t} k_0 \exp\biggl ( -\frac{\Delta E}{k_B\tilde{T}(0,t)}\biggr )\,
  dt=\Delta t k_0 \exp\biggl ( -\frac{\Delta E}{k_BT}\biggr )\,.
\end{equation}
We have neglected here the temperature dependence of $k_0$. When
including it, we found that the resulting effective temperature
changes only slightly. We have chosen $\Delta t$, as the average time
between two ions arriving in a circle with area $\pi r_0^2$
under the ion beam spot,  resulting in $\Delta t=1.4\times 10^5$ps.
For the energy barrier, we have chosen $\Delta
E=E_{SB}+3E_{NN}+E_S$, which corresponds to atoms along edges of
islands/steps. Using these parameters, we found an effective surface
temperature of $T=1200$K, i.e. considerably higher than room
temperature. In this calculation it is assumed that only the energy
carried by the ions hitting the ``target area'' $\pi r_0^2$ contribute to
the heating of the surface inside the area. 
If one takes into account that also the
ions hitting the neighborhood of the target area contribute to the
heating inside the area, even higher effective 
temperatures can expected.

\begin{figure}[!htbp]
\begin{center}
\includegraphics[angle=270, width=0.45\textwidth]{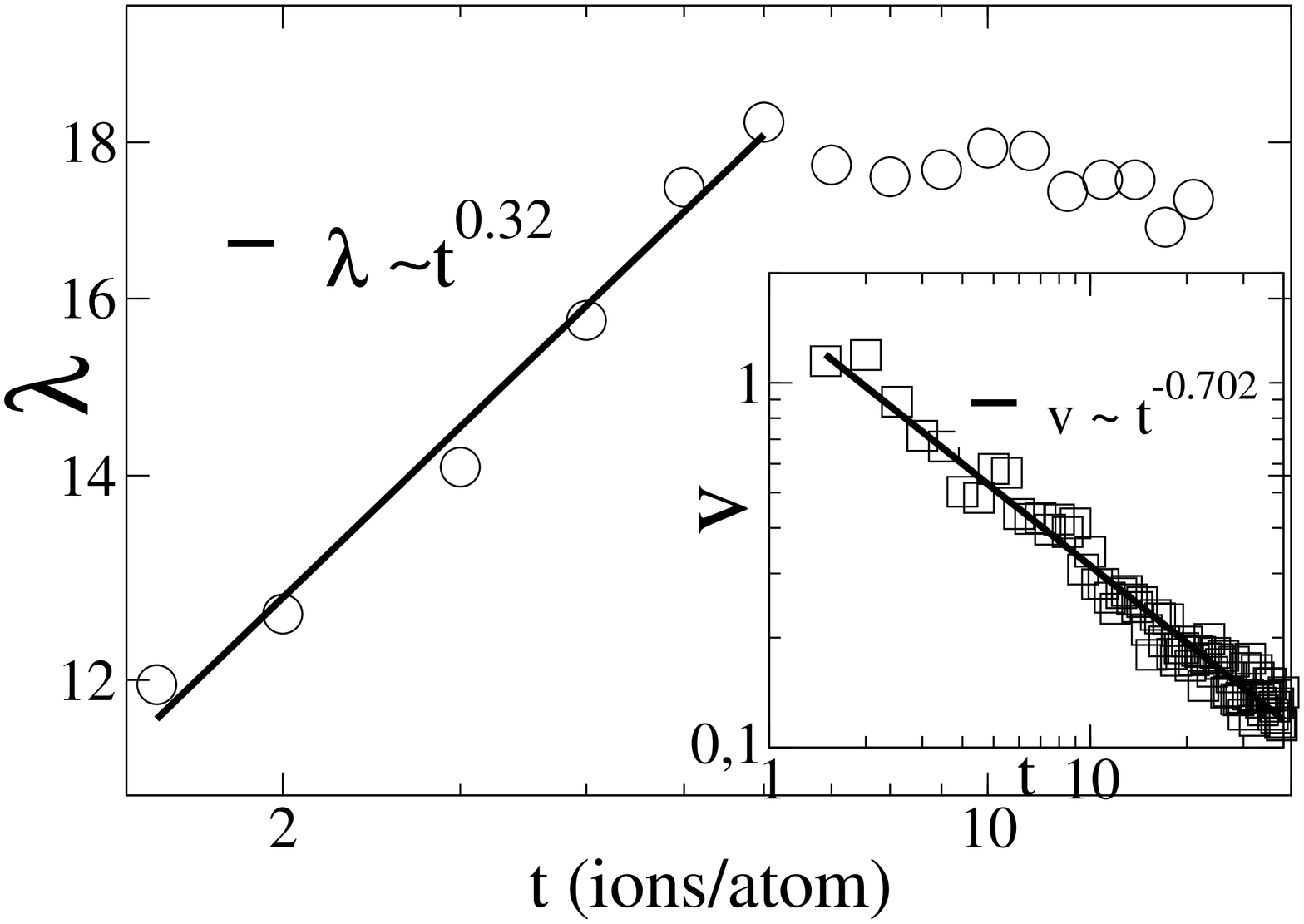}
\caption{\label{fig:wavelength_tasd0.1} 
Ripple wavelength, $\lambda$, measured in lattice units, as a fuction of
  time, $t$. The inset shows the time dependence of the ripple propagation
  velocity, $v$ (measured in lattice units per ion per atom). Both
  results are for the kinetic diffusion mechanism, at a substrate
  temperature of $k_BT= 0.1 eV$.}
\end{center}
\end{figure}

The exact effective temperature depends on many parameters as
ion energy, ion current density, heat conduction, surface roughness etc. We are
here interested only in universal effects, not in modeling a specific
experimental setup. For this reason, we use the above
result only as a guideline and study several temperatures of this
order of magnitude and additionally above. 
Hence, for the further analysis of ripple movement, we consider high effective
temperatures for the Arrhenius MC model, such that
the surface diffusion is indeed able to
 act as an effective smoothing mechanism (see
Fig. \ref{fig:profile2}). At such higher temperatures we observe
some universal features for both diffusion mechanisms, as presented
now. Figure \ref{fig:wavelength_tasd0.1} is the plot of the ripple
wavelength (circle symbols) versus time measured in units of the
number of ions per atom; its inset is a plot of the projection of the
ripple velocity along the ion beam direction, 
versus time, both at the estimated effective temperature of $k_BT= 0.1
eV$ corresponding to the experimental conditions from Ref.\
\cite{habenicht2002}.

 A plot of wavelength versus
time in Fig. \ref{fig:wavelength_tasd0.1} reveals that for short times
$\lambda \sim
t^{0.32}$, which is in-between the results $\lambda \sim
t^{0.5}$ of Habenicht et al. and $\lambda \sim t^{0.26}$ of Frost et
al. \cite{Frost2000}. But we observed a power-law behavior only in the
initial stages of ripple formation, the wavelength becoming constant
in time at the later stage.

The velocity shows a  power-law behavior over a larger time interval, 
resulting in $v \sim t^{-0.7}$ as obtained from inset of
Fig. \ref{fig:wavelength_tasd0.1}. This is in excellent agreement with the
experimental result $v \sim t^{-0.75}$ of Habenicht et al. 
\cite{habenicht2002}. A difference is that for smaller times a
constant velocity was observed in the experiments, while we do not see
any clean ripples for smaller times than the power-law regime. Anyway,
combining both scaling results
gives $v \sim \lambda^{-2.19}$, in good agreement with the exponent $-2$ of
continuum theory \cite{makeev2002}. 

\begin{figure}[!htbp]
\begin{center}
\includegraphics[angle=270, width=0.45\textwidth]{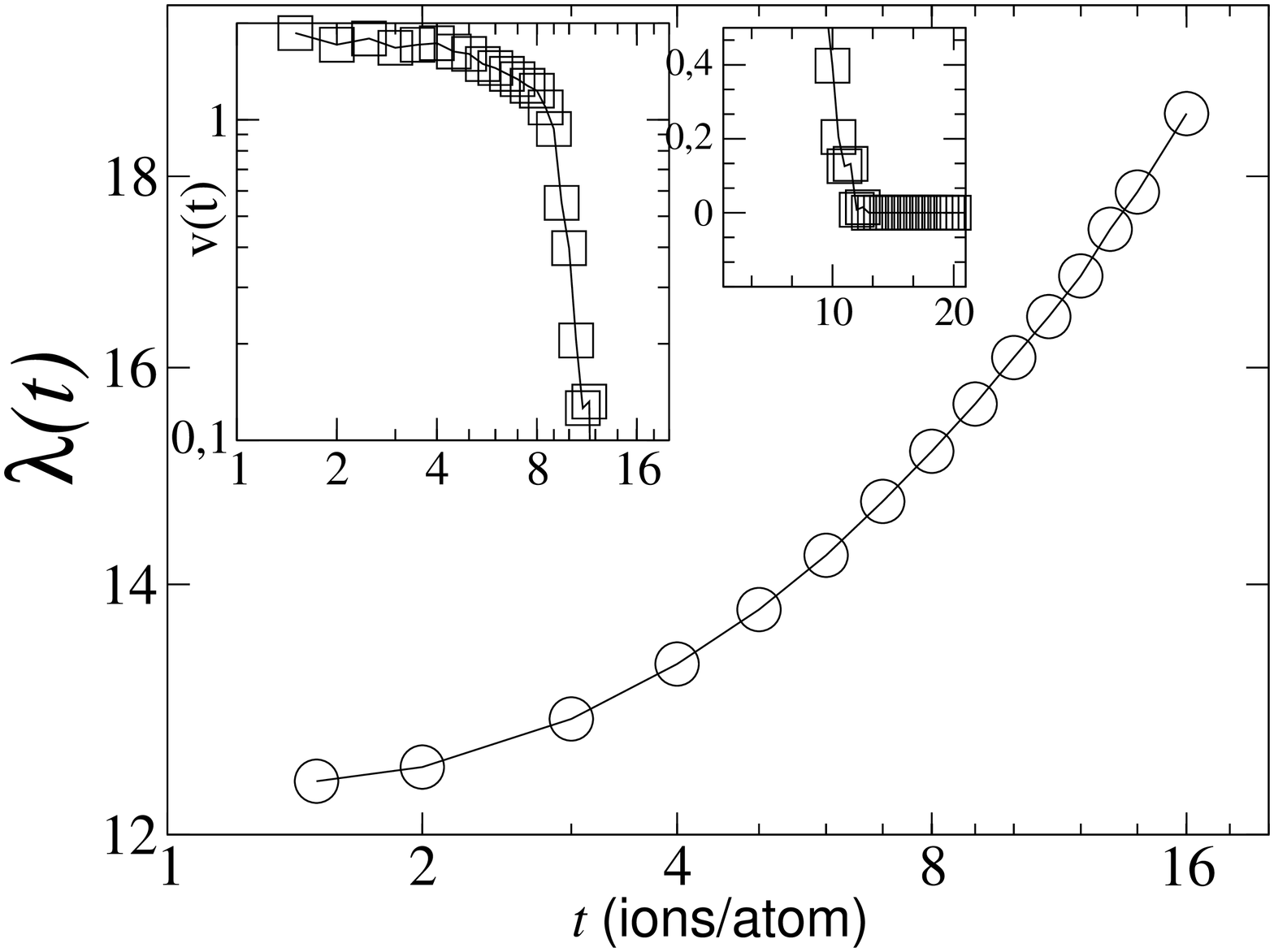}
\caption{\label{fig:wavelength_asd} 
Ripple wavelength, $\lambda$, measured in lattice units, as a fuction of
  time, $t$. The inset shows the time dependence of the ripple propagation
  velocity, $v$ (measured in lattice units per ion per atom). Both
  results are for the thermodynamic diffusion 
  mechanism, at a substrate
  temperature of 0.2 $Jk_B^{-1}$.}
\end{center}
\end{figure}
\begin{figure}[!htbp]
\begin{center}
\includegraphics[angle=270, width=0.45\textwidth]{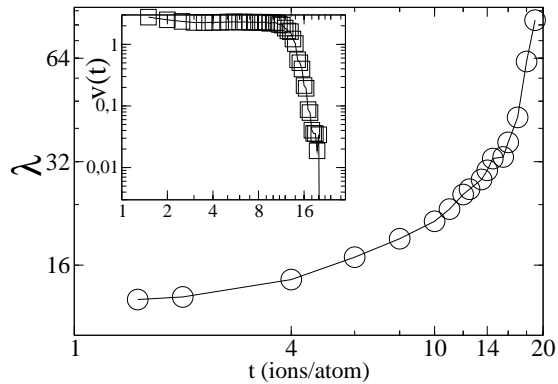}
\caption{\label{fig:wavelength_tasd} Same plot as in
  Fig. \ref{fig:wavelength_asd}
 but for the Arrhenius diffusion mechanism, 
  for a substrate temperature of 0.2 $eV/k_B$. In both figures, the
line with circle symbols represent the wavelength while the line with
square symbols represent the velocity.}
\end{center}
\end{figure}

Now we turn to higher effective subtrate temperatures, corresponding
e.g.\ to higher ion currents and/or materials with lower heat-conductivity.
Figures \ref{fig:wavelength_asd} and \ref{fig:wavelength_tasd}
 are plots of the ripple wavelength (circle symbols)
as a function of time, at respective temperatures $k_BT= 0.2 J$ and
$k_BT= 0.2 eV$; using the first 
and second models of surface diffusion respectively. 
{In both models, the ripples disappear after a while, i.e.\ the ripple
wavelength diverges.}
Considering
the lifetime of the ripples from first appearance to 
annihilation, the wavelength
increases exponentially with time as $\lambda \sim \exp(\rho t)$,
$\rho=0.029$ (Fig.\ref{fig:wavelength_asd} ), 
in the first model, while it increases with
time according to the 
inverse law $\lambda(t)\sim 1/(c_1-c_2t)$ with $c_1=0.083$ and
$c2=0.0036$ (Fig. \ref{fig:wavelength_tasd}) 
in the second model. 
To investigate the origin of the difference, we performed also
simulations with the Arrhenius model, but with the Schwoebel term set to
zero. In this case the result was very similar to result in
Fig. \ref{fig:wavelength_asd} of the thermodynamic
model (which has no Schwoebel term here), 
and we obtained a behavior $\lambda \sim \exp(0.036t)$.
On the other hand, 
when we set the energy in the Schwoebel term to twice its value,
$E_{SB}=0.3$eV, the result is very similar to $E_{SB}=0.15$eV. This
shows that the Schwoebel barrier plays an important role in the pattern
formation process.
Note that these fits are purely heuristic. We are
not aware of any theory of the time dependence of ripple wavelength
and velocity, only a calculation of the dispersion relation
$v(\lambda)$ has been performed within linear theory
\cite{makeev2002}. Furthermore, there exists an analytic  
study of the temporal development of step
bunches during epitaxial growth \cite{pimpinelli2002,remark_pimpinelli}.

\begin{figure}[!htbp]
\begin{center}
\includegraphics[angle=270,
width=0.5\textwidth]{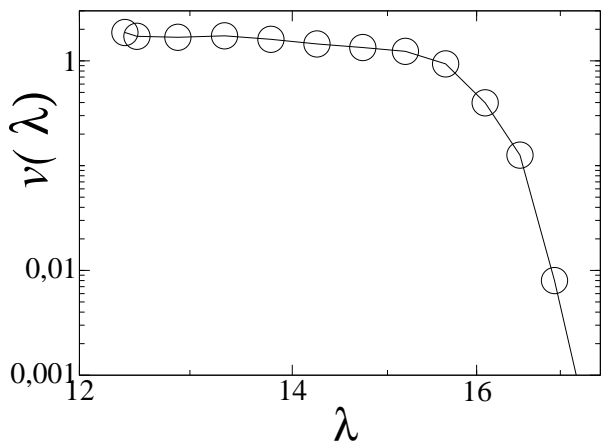} 
\caption{\label{fig:velocity} Ripple velocity as a function of ripple
  wavelength, for the thermodyanmic and, in the inset, for the Arrhenius
  surface diffusion mechanism.}
\end{center}
\end{figure}

\begin{figure}[!htbp]
\begin{center}
\includegraphics[angle=270,
width=0.5\textwidth]{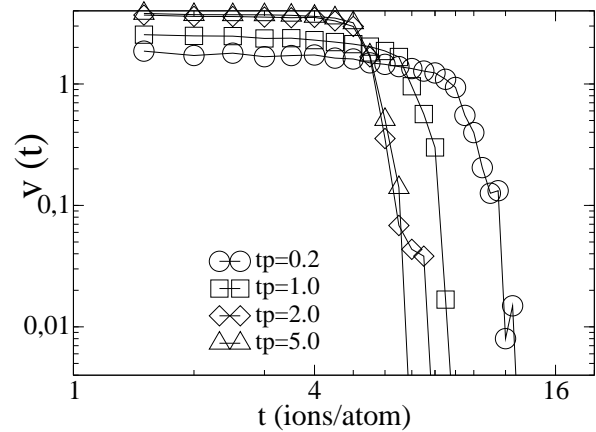} 
\includegraphics[angle=270,
width=0.5\textwidth]{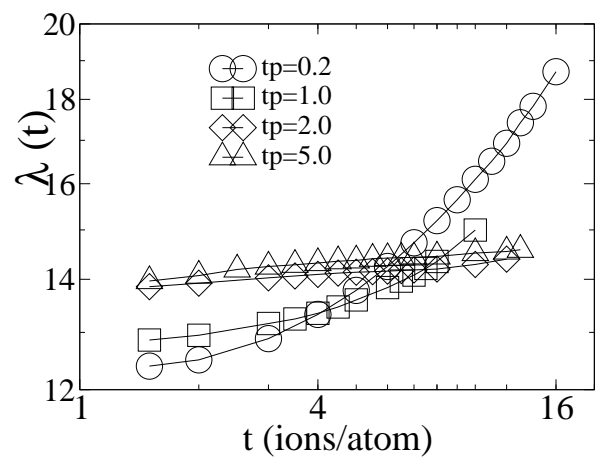} 
\caption{\label{fig:temperature} Temperature dependence of the ripple
  velocity (upper graph), and ripple wavelength (lower graph); for the
  thermodynamic diffusion mechanism. Temperature is in units of 
  $Jk_B^{-1}$.} 
\end{center}
\end{figure}

The insets of Figs. \ref{fig:wavelength_asd} and
\ref{fig:wavelength_tasd} 
are plots of the ripple velocity (line
with square symbols) as 
a function of time. Irrespective of which surface diffusion mechanism
is employed, the velocity is at first almost independent of
time, then it disperses after a transition time $t_r$. 
{This initial plateau is similar to the plateau observed in the
experiments, but the drop in velocity is very abrupt, no clear power
law is visible then.}
Moreover, the ripples finally come to rest before completely
disappearing, as seen in the smaller inset of
Fig. \ref{fig:wavelength_asd}. We find, however, that at the lower
temperature in the kinetic model, the ripples do not
stop moving until their disappearance.  
Figure \ref{fig:velocity} shows the dependence of the ripple velocity on
the wavelength 
{for $k_BT=0.2$eV resp $0.2J$}, 
their order of magnitude relationship is about the
same as in the experiment. 
We see in Fig. \ref{fig:temperature} that the trend in
velocity  
variation is the same at high temperatures but the magnitude
increases with temperature, as one would expect from the temperature
dependence of the surface diffusion. 
But we only observed a power-law scaling at temperatures below
$k_BT\approx 0.18 eV$. This
indicates that the presence of  power-law scaling of ripple wavelength
and velocity, and the corresponding exponents, 
depend on the time scale of observation
(Fig. \ref{fig:wavelength_tasd0.1}), as well as on the 
effective temperature.

It seems that the
increase in magnitude of the velocity, when measured 
at same time ($t<t_r$) but different
temperatures,  does not continue indefinitely in our model. In the
upper graph of Fig. \ref{fig:temperature}
there is very little difference in the
magnitudes of the velocity at temperatures $2.0$ and $5.0Jk_B^{-1}$;
even though the temperature difference is very high. 
This saturation behaviour is also displayed
in the ripple wavelength at the same higher temperatures, as seen in
the lower graph of Fig.  \ref{fig:temperature}. 
In principle, one can still fit an  exponential law to the
data, except that the decay constant $\rho$ in the exponential 
becomes very small (It has the respective values of $0.029$, $0.018$, $0.0031$, and
$0.003$ from the lowest to the highest temperature.). So for very 
high effective temperatures we could
equally well fit a power-law. Hence, there may be
some ``critical substrate temperature'', above which the wavelength remains
nearly constant in time; and the velocity, after some time $t_r$,
drops instantaneously to zero. Nevertheless, the temperature where
such a ``transition'' will take place, is probably unphysically high
(see below),  so that the material 
{ used in the experiment
would start to evaporate before reaching this point. But other materials,
in combination with high ion currents, might quite display such a behavior.}
So far, we are only aware of one set of experiments 
\cite{habenicht2002}. Hence, it would be very
interesting to see, whether some temperature dependence of the
dynamical features, including the disappearance of the coarsening, can
be seen in experiments at higher effective temperatures corresponding
to high ion currents and/or higher lab temperatures.

Our results for the second diffusion model
also indicate that in $\lambda(t)\approx1/(c_1-c_2t)$, $c_2$ approaches
zero with increasing temperature. Here, where we can measure the
temperature in real units, it is clear that the ``transition'' to
almost non-coarsening ripples, takes place at unrealistically high
temperatures $2-5$eV/$k_B$, where the material starts to evaporate.
Moreover, we notice in Fig.  \ref{fig:temperature} 
that the transition time from
linear regime to onset of non-linearities decreases with
increasing temperature. 

To summarize, 
ripple propagation depends on the effective substrate temperature as
well as diffusion mechanism. At around so-far experimentally
realized temperatures, ripples
propagate, from first appearance, with decreasing velocity until
disappearance without full cessation of motion. At high effective temperatures,
however, immediately after ripple formation, the ripples
move with constant velocity for some time, after which they begin to 
decelerate (insets of Figs. \ref{fig:wavelength_asd},  
\ref{fig:wavelength_tasd})  and after some time, depending on the diffusion
model, the ripples stop moving but keep disappearing gradually.
 At the same time the ripple structure is
gradually being washed out, and in the final stage the ripples
are completely wiped out. The ripple wavelength is always increasing
in time at high temperatures, while at low effective temperatures it initially
increases with time, and later becomes constant.  

\section{\label{sec:conclusion}Conclusion and Outlook}

We have studied the propagation of ripples by means of
a discrete ($2+1$)-dimensional model of the sputtering process,
combined  with one of two different solid-on-solid
models of surface diffusion: a Arrhenius MC mechanism with Ehrlich-Schwoebel
barriers and a thermodynamic mechanism without a Schwoebel term. 
We have obtained the formation and propagation of the
ripples with both diffusion mechanisms used in turn. Furthermore, we
have obtained 
the same trend in the behavior of ripple velocity and wavelength as
observed experimentally and predicted theoretically, but, in addition
to the experimental results, we  
find a drastic change in the ripple propagation at temperatures well
above the so-far experimentally realized effective
 temperature; for instance we found deviations from
power-law into exponential or inverse-law behavior, and in addition,
the ripples first stop moving before vanishing 
completely. We find that, at very 
high effective temperatures, the behavior of the
ripple velocity is 
charaterized by two regions, separated at the transition time. In the
first region it is constant and in the second region it decreases
rapidly to zero. Between the two regions a power-law dependence can be
observed for 
some small time interval. Whereas, around so-far 
experimentally realized  temperature, the
velocity-time relationship obeys a power-law.
Furthermore, at high effective temperatures, the wavelength increases
exponentially with time in the thermodynamic diffusion model 
(and in the Arrhenius diffusion model without Schwoebel term)
and
obeys an inverse law for the Arrhenius model including the Schwoebel
barrier. In addition,
we find further strong dependencies on the effective substrate
temperature; as the temperature increases the magnitude of the
velocity also increases. 
The transition time between constant and decreasing velocity is also found to
decrease with increasing temperature. Our results indicate an
approach towards a saturation behaviour of velocity or wavelength with
increasing effective substrate
temperature, where the wavelength is expected to become time
independent. However,  this may happen at an unphysically
high temperature. Anyway, an experimental
study of the dependance of the dynamical features of
ripple formation and effacement on the physical conditions seems very
promising.

One open problem of our model, at high incidence angles
(e.g $\theta > 75^\circ$),  is that it uses the Sigmund formula for
modeling the sputtering process. In a recent simulation
\cite{feix2004}  using a binary collision approximation, we observed
that close to the penetration point of the ion, much less atoms were
sputtered than predicted by the Sigmund formula (\ref{eq:sigmund}), in
fact the distribution shows a  minimum there. When incorporating this
effect in the Bradley-Harper linear theory \cite{bradley1988} of
sputtering, one e.g. observes \cite{feix2004}
that the sputter yield, i.e. the number
of removed atoms per ion, exhibits a minimum for grazing incidence,
like in experiments, in contrast to the orginial linear theory
\cite{bradley1988}. Hence, it may be promising to apply a different
formula describing the sputtering, which takes into account this
effect.  

Furthermore, the role of the interplay between the surface diffusion
process and the sputtering process is still not fully
understood. So far, we know that including a pure $T=0$ relaxation in
our sputtering model does not \cite{hartmann2002} lead to a
disappearance of ripples for long times. Next, we know from this study,
that one approach including calibrated Schwoebel 
barriers does not yield ripples at
room temperature for sputtering yield $Y\approx 7$. There are several
different models \cite{das-sarma1991,wilby1992,Siegert1994,Smilauer1993,ramana-murty1999, 
malarz1999,ghaisas2001,pundyindu2001} 
for surface diffusion, which could be combined in a construction kit
manner. Here, an extensive study over different combinations of
parameters is necessary.

Finally,  
it would be of interest to include crystal anisotropy into the surface
diffusion; This may give results in
agreement with experimental studies of metallic substrates, which may
be useful in understanding the anomalies of such surfaces.

Acknowledgements: 
The authors would like to thank K. Lieb and R. Cuerno for 
helpful discussions and suggestions. 
OEY thanks Henning L\"owe for interesting discussions. 
The large scale numerical simulations were performed on the
workstation clusters of the institute. 
This work was funded by the DFG
({\em Deutsche Forschungsgemeinschaft}) 
within the SFB (Sonderforschungbereich) 602 and by
the {\em VolkswagenStiftung} (Germany) within the program
``Nachwuchsgruppen an Universit\"aten''.

\end{document}